\documentclass[reprint,amsmath,amssymb,aps,pre,showpacs]{revtex4-1}
\usepackage{amssymb, amsmath, amsbsy}
\usepackage{graphicx}
\usepackage{color}
\begin{document}

\title{Anisotropy of flow in stochastically generated porous media}
\author{Maciej Matyka}
\author{Zbigniew Koza}
\author{Jaros{\l}aw Go{\l}embiewski}
\affiliation{Faculty of Physics and Astronomy, University of Wroc{\l}aw, 50-204 Wroc{\l}aw,
Poland}
\author{Marcin Kostur}
\author{Micha{\l} Januszewski}
\affiliation{Institute of Physics, University of Silesia, 40-007 Katowice, Poland}

\date{\today}

\begin{abstract}
Models of porous media are often applied to relatively small systems,
which leads not only to system-size-dependent results, but also to phenomena that would be absent in larger systems.
Here we investigate one such finite-size effect: anisotropy of the permeability tensor.
We show that a non-zero angle between the external
body force and macroscopic flux vector exists in three-dimensional
periodic models of sizes commonly used in computer simulations
and propose a criterion, based on the system size to the grain size ratio, for this phenomenon to be relevant or negligible.
The finite-size anisotropy of the porous matrix induces a
pressure gradient perpendicular to the axis of a porous duct and we analyze how this effect
scales with the system and grain sizes.
\end{abstract}

% 47.15.G- == Low-Reynolds-number (creeping) flows
% 47.56.+r == Flows through porous media
% 91.60.Np == Permeability and porosity

\maketitle

\section{Introduction\label{sec:intro}}

For technical reasons pore-scale simulations of flow through porous
media are
based on simplified, finite-size models.
For example, it is not unusual for a  numerical model of a bed of sand to contain only
8  ``sand grains'' \cite{Sugita12},
and typical values of the ratio
of the system size  to the grain diameter, $L/a$, used in numerical simulations of three-dimensional (3D)
flows range from $ \approx 31$ \cite{Pan01} through $17$ \cite{Knackstedt94,Zhang95}, $10$
\cite{Yiotis07}, $8$ \cite{Succi89,Cancelliere90}, $3$
\cite{Verberg99} down to only $  2$ \cite{Sugita12}.
Numerical systems with
a small value of $L/a$ are likely to be smaller than
the representative elementary volume (REV) of the porous medium they
model \cite{Zhang00}.
Consequently, results of many numerical simulations
can contain some systematic errors if this fact has not been properly accounted for.

The basic macroscopic law for creeping flow through a porous medium is
Darcy's law,
\begin{equation}
 \label{eq:Darcy}
   \mathbf{q} = -\hat{\mathbf{K}}(\nabla P - \mathbf{\rho g} )
\end{equation}
where $\mathbf{q}$ is the volumetric fluid flux,
$\hat{\mathbf{K}}$  is a symmetric tensor of the hydraulic
conductivity,
$\rho$ is the fluid density,
$\mathbf{g}$ is an external bulk force per unit mass, e.g., gravity, and $\nabla P$
is the pressure gradient \cite{Bear72}.
If the system under consideration is smaller than the REV, $\hat{\mathbf{K}}$ becomes dependent
on a particular realisation of the porous medium, and its components can be regarded as random variables.
In particular, a small-size sample of an isotropic medium can exhibit quite high anisotropy.
Analysis of the anisotropy of $\hat{\mathbf{K}}$ as a function of the system size can thus serve
as a way to estimate the size of the REV in computer simulations or in some small-scale experiments,
e.g., particle image velocimetry \cite{Morad09}.

The finite-size anisotropy of a statistically uniform porous medium can be estimated
by the dispersion $\sigma_\alpha$ of the angle $\alpha$  between the volumetric fluid flux ($\mathbf{q}$)
and the driving force ($\mathbf{\rho g} - \nabla P$) about its mean value.
Using some phenomonological  arguments, we proposed \cite{Koza09} a general relation
\begin{equation}
 \label{eq:delta_d}
  \sigma_\alpha \propto L^{-\delta},\quad \delta = d/2,
\end{equation}
where $d$ denotes the space dimension and $L$ is the system size,
and verified it numerically for two-dimensional (2D) and quasi one-dimensional models.
The aim of the present  study is to investigate finite-size anisotropy of statistically uniform porous media
in the more realistic three-dimensional case.

\section{Model\label{sec:model}}

\begin{figure}
\centering
\includegraphics[width=0.7\columnwidth]{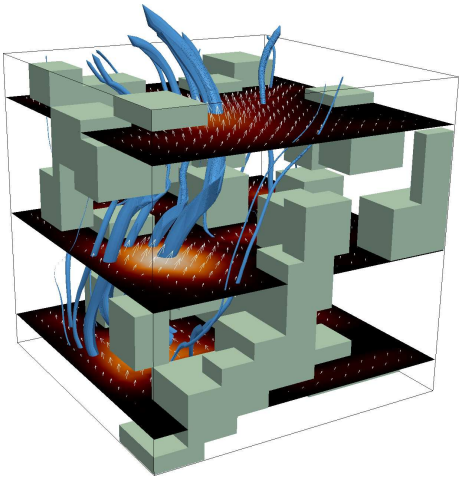}
\caption{(Color online)  Exemplary realisation of a porous matrix
and the corresponding flow solution.
The matrix is composed of overlapping solid cubes
and periodic boundary conditions are assumed in all directions.
Small arrows
show the velocity field and
its magnitude is represented by
the thickness of streamtubes and the brightness of the color on the horizontal cross-sections.
The porosity $\varphi=0.7$.
\label{fig:definition}
}
\end{figure}

As visualised in Fig.\ \ref{fig:definition},
we model the porous matrix as a system of identical,
impenetrable cubes of side $a$ randomly distributed in a cuboidal duct.
The cubes are aligned along the edges of the system, they are free to overlap and their number
depends on the target porosity.
The system is always periodic along the duct axis
and in the remaining two directions
either periodic boundary conditions or no-slip, impenetrable walls are assumed.
Due to periodicity, the pressure gradient along the duct axis vanishes.
The fluid is set in motion by a body force directed along the
duct axis. The fluid is also assumed to be incompressible and its
velocity is small enough to ensure that the flow is in the Darcy regime.
Models of this type were studied, for example, in Refs.
\cite{Morais09,Tomadakis93,Koponen97,Matyka08} and more recently in \cite{Duda11}.

\section{Simulation methods\label{sec:simul-method}}

We chose  the lattice Boltzmann method (LBM) as the main simulation method.
In this approach, the simulation domain is discretized into a regular lattice and
the fluid is described by a mesoscopic particle distribution function
 defined on all nodes of the lattice. The mesoscopic velocity is also discretized, allowing momentum
and mass transfer between selected neighboring nodes only. In this work, we use the
D3Q19 lattice
with the so-called multi-relaxation times (MRT) collision operator \cite{Dhumieres02}.
For non-slip boundary conditions, we use the bounce-back rule
and body forces are implemented using Guo's method \cite{Guo02}.
This method
can be shown to reproduce the Navier-Stokes equations with second-order accuracy.
 We refer the reader
to Refs.\ \cite{Chen98, Aidun10} for a more detailed discussion of the LBM.

Of many implementations of the LBM we chose
the Sailfish library (http://sailfish.us.edu.pl).
A unique feature of Sailfish is that it runs on graphics processing units (GPUs),  which
allows for significant speedups as compared to typical CPU codes~\cite{tolke-GPU}.
Sailfish automatically
generates optimized GPU code
based on a high-level model description consisting
in a large part of formulas in a computer algebra system (Sympy).
The code takes advantage of the
intrinsic parallelism of the lattice Boltzmann method by assigning individual lattice
nodes to separate threads on the GPU. These threads then execute almost independently,
only exchanging data with nearest neighbors in the streaming step of the LB algorithm.
The details of the implementation and code optimization techniques exploited in the library will be presented
elsewhere.

One of the main drawbacks of the LBM is that it is based on a regular lattice.
This means that a porous matrix is actually approximated by a set of  small cubes (mesh cells)
and may contain many artificial edges and vertices.
This, in turn, introduces relatively large errors into the flow solution at nearby mesh cells.
A solution obtained on a regular lattice can be thus regarded either as an approximate solution
for given boundary conditions or as a valid solution for a model where only an approximate location
of the boundaries is known. While in principle this uncertainty could be reduced by increasing the mesh resolution,
this approach could quickly exhaust all available computer resources.

Instead, we verified the LBM solutions by the finite volume method (FVM).
In this approach the space is divided into a mesh of polyhedrons which are then used, through volume integrals,
to express the Navier-Stokes equations in the form of algebraic equations.
Meshing was performed with the snappyHexMesh (SHM) utility from the OpenFOAM toolkit \cite{OpenFOAM}.
Due to hardware limitations, the number of cells was limited to $\approx$ 2 million.
The flow equations were solved with OpenFOAM's icoFoam solver, which is an open-source
implementation of the PISO algorithm \cite{Ferziger-Peric},
a well-established method of computational fluid dynamics.
The code of icoFoam was modified to adjust it to the needs of our simulations,
e.g.\ handling of the external force $\mathbf{g}$.
While PISO allows for much larger time steps than the LBM and can be run on unstructured meshes
adjusted to the local geometry of the porous medium, it turned out far more time- and resource-consuming,
therefore we used it only for verification purposes on relatively small systems.

By varying the magnitude of driving force we checked that the flow is well within the Darcy regime.
To ensure that the steady-state had actually been reached, for systems with periodic
boundary conditions
we monitored the temporal evolution of the angle $\alpha$ between the external body force $\mathbf{g}$
and the resulting Darcy flux $\mathbf{q}$ and continued the simulations until
\begin{equation}\label{eq:steady-state}
   |\alpha(k) - \alpha(k - N)| \le \varepsilon|\alpha(k - N)|,
\end{equation}
where $k$ is the LBM or PISO iteration number, $\varepsilon=10^{-6}$,
and $N=2\times 10^4$ or $N = 500$ for the LBM and PISO, respectively.
In the case of a solid wall channel we used an alternative stopping criterion,
\begin{equation}
\sqrt{ \frac{ \sum_{i}\left|\mathbf{u}_{i}(k) -
\mathbf{u}_{i}(k-N)\right|^2 }{\sum_{i}|\mathbf{u}_{i}(k)|^2} } \le \varepsilon.
\end{equation}
where $\mathbf{u}_i$ is the local velocity at cell $i$.

To verify our computational procedures, we applied both the LBM and PISO to determine
the permeability of a simple cubic arrangement of overlapping
spheres. This standard problem was already investigated with several independent numerical methods
and reference solutions are available for comparison \cite{Larson88,Holmes11}.
We found the permeabilities computed with both methods
to be in acceptable agreement (a few percent) with the reference values (data not shown).

\section{Results\label{sec:results}}

\subsection{Periodic boundary conditions \label{ssec:PBC}}

If the system has periodic boundary conditions in all directions, the
angle $\alpha$ between the body force $\mathbf{g}$
and the Darcy flux $\mathbf{q}$ is usually nonzero and
its standard deviation, $\sigma_\alpha$, taken over an ensemble of porous samples,
is expected to depend on the system size $L$ in accordance with
Eq.~(\ref{eq:delta_d}). To verify whether this equation
holds for 3D systems,
we generated hundreds of independent
porous samples, as described in Sec.~\ref{sec:model}, with $a=4$ lattice units (l.u.).
We chose two target porosities, $\varphi = 0.35,0.7$, lying between the  porosity
below which the flow is completely blocked, $\varphi_\mathrm{b}\approx 0.1$, and the percolation threshold
(for overlapping cubes) above which the porous matrix splits into disjoint clusters,  $\varphi_\mathrm{p} \approx 0.77$.
Simulations were performed for several system sizes $L$ ranging from 10 to $260$ l.u.\
and the results were averaged over 25 independent samples.
The number of obstacles per sample varied from $\approx 6$ ($L/a = 2.5$, $\varphi \approx 0.7$)
to
$\approx 284~000$ ($L/a = 65$, $\varphi \approx 0.35$). The actual porosity 
of a random sample could differ from the target porosity by $\approx 5\%$ for $L/a = 2.5$ 
and by  $\approx 0.03\%$ for  $L/a = 65$.

\begin{figure}
\centering
\includegraphics[width=0.9\columnwidth]{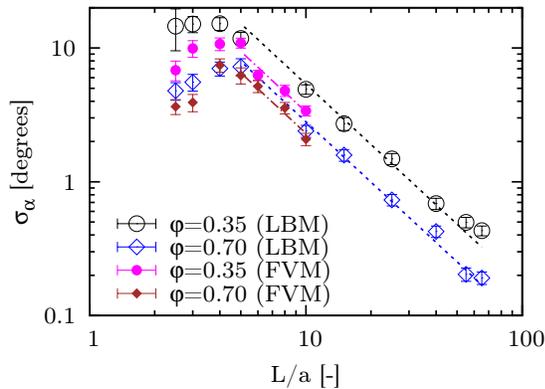}
\caption{(Color online)
Dispersion of the angle between the flow direction ($\mathbf{q}$)
and the external force
($\mathbf{g}$) as a function of the system size $L$ for two porosities
$\varphi$ and $a=4$ l.u.
Open and filled symbols correspond to the LBM and FVM (PISO) solvers, respectively.
Dashed lines are the best fits to the theoretical formula
$\sigma_\alpha \propto (L/a)^{-3/2}$ for $L/a\ge 5$.
Error bars represent the standard errors of the mean
calculated from $25$ independent samples.
\label{fig:sigmal}}
\end{figure}

The results, shown in Fig.~\ref{fig:sigmal},
are in good agreement with Eq.~(\ref{eq:delta_d}).
It is interesting to notice that this asymptotic formula describes
the system behavior even for as small values of $L/a$ as 5.
As expected \cite{Koza09}, the finite-size anisotropy increases as the porosity is decreased,
which can be related to the fact that the flow in low-porosity media is more tortuous,
a factor particularly important close to the percolation threshold.

Comparison of the results obtained with the PISO and the LBM reveals that the former tends to
give systematically smaller values of $\sigma_\alpha$.
Nevertheless, both methods yield the same value of   $\delta\approx 3/2$ in Eq.~(\ref{eq:delta_d}).
We checked that the discrepancy between the results could be significantly
reduced by using finer meshes, at the cost of one being forced to consider smaller system sizes due to hardware limits
(data not shown).
Because of complex meshing,
PISO turned out to be particularly  costly in terms of computer
time and resources.
For example, a single PISO simulation for the case of $L/a=10$ took $\approx$ 7 hours (including the meshing)
and required 4 GB of RAM on a single core of an AMD Opteron 2384 processor,
whereas the Sailfish code running on a Tesla M2090 GPU
would solve the same case in less than $5$ minutes, requiring
only about 12 MB of storage.
For this reason we did not use PISO for systems with $L/a>10$.
Detailed comparison of the two methods in the context of porous media studies will be discussed elsewhere.

\subsection{No-slip boundary conditions}

If a porous medium is placed inside a long, straight  channel with impermeable walls
then the mean fluid velocity will be parallel
to the channel boundaries irrespective of the direction of the driving force $\mathbf{g}$.
If the medium is anisotropic
then a pressure gradient
$\nabla_{\!\perp} P$ perpendicular to the channel axis
must develop for Darcy's law, Eq.~(\ref{eq:Darcy}), to be satisfied.

\begin{figure}
\centering
\includegraphics[width=0.9\columnwidth]{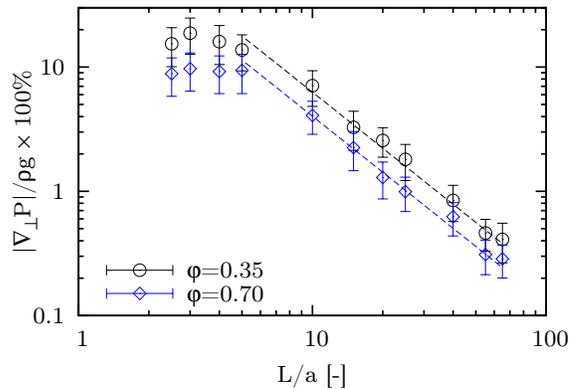}
\caption{ (Color online) LBM results for the magnitude of the induced pressure gradient,
 $|\nabla_{\!\perp} P|$,  normalized by the  hydrostatic pressure gradient, $\rho g$, averaged
 over $25$ configurations.
 The dashed lines are the best fits to $|\nabla_{\!\perp} P|/\rho g \propto (L/a)^{-3/2}$ for $L/a\ge 5$.
 \label{fig:pressure_grad}
}
\end{figure}

To investigate this effect we used the same model as in
Sec.~\ref{ssec:PBC} except for the no-slip boundary conditions imposed
on the channel walls.
The results, obtained with the LBM for several values of $\varphi$ and  $L/a$
and normalized by the gradient of the hydrostatic pressure acting along the channel,
are depicted in Fig.~\ref{fig:pressure_grad}.
As expected, the dependence of $|\nabla_{\!\perp} P|$ on the system size turns out similar to that of
$\sigma_\alpha$ in periodic systems, i.e., $|\nabla_{\!\perp} P| \propto L^{-3/2}$ for $L/a \gtrsim 5$.
This result can be generalized to systems
of arbitrary space dimension $d$ and expressed in a dimensionless form as follows,
\begin{equation}
 \label{eq:nabla-P}
  \frac{|\nabla_{\!\perp} P|}{\rho g} \propto \left(\frac{L}{a}\right)^{-d/2},  \quad \mbox{for } L/a \gg 1.
\end{equation}

\subsection{Condition for the induced pressure gradient to vanish\label{ssec:no_grad_flow}}
Equation~(\ref{eq:Darcy}) implies that it is possible to eliminate the induced pressure gradient
by replacing  $\mathbf{g}$ with
\begin{equation}
 \label{eq:g'}
  \mathbf{g}' =  \mathbf{g} + \nabla_{\!\perp} P(\mathbf{g})/\rho
\end{equation}
where
$\nabla_{\!\perp} P(\mathbf{g})$ denotes the pressure gradient induced by $\mathbf{g}$.
Let $\beta$ denote the angle between $\mathbf{g}'$ and the
channel axis. Equation~(\ref{eq:nabla-P}) can be now rewritten as
\begin{equation}
 \label{eq:beta-tan}
  \tan \beta \propto \left(\frac{L}{a}\right)^{-\delta}.
\end{equation}
As $\beta$ is small, this formula can be further simplified to
\begin{equation}
 \label{eq:beta}
   \beta \propto \left(\frac{L}{a}\right)^{-\delta}.
\end{equation}
This means that $\beta$ decreases with an increasing system size in the same way as
$\sigma_\alpha$ does, cf.\ Eq.~(\ref{eq:delta_d}).

\subsection{The size of the REV}
Generally, the size of the REV depends on many factors, including the required accuracy of measurements.
Its magnitude can be estimated in our model based
on the dependence of $\alpha$ or $|\nabla_{\!\perp} P|/{\rho g}$ on the system size.
If we make an \emph{ad hoc} assumption that the anisotropy effects are small enough to be practically negligible
for $\sigma_\alpha \lesssim 2^\circ$
then
from Fig.~\ref{fig:sigmal} one can estimate the size of the REV to be $\approx 20a$.
This value agrees with the data in Fig.~\ref{fig:pressure_grad}, which shows that
for $L/a>20$ the induced pressure gradient is smaller than 2\%  of the
pressure gradient driving the fluid through the duct. This enables to estimate
the size of the REV in the model considered here as $\approx 80\times 80\times 80$ l.u.
Note that since the finite-size anisotropy effects grow as the porosity is lowered,
cf.\ Figs.\ \ref{fig:sigmal} and \ref{fig:pressure_grad},
the size of the REV is also expected to increase as $\varphi$ is decreased.
For porosities close to $\varphi_\textrm{b}$, which is the percolation threshold for the
pore space, the size of the REV is expected to grow to infinity.

\section{Discussion of the results and conclusions\label{sec:conclusions}}
Pore-scale computer simulations of transport through porous media are very resource-intensive and hence
are often performed on systems whose size is smaller than the representative elementary volume,
which makes the results system-size-dependent.  In particular, statistically generated homogeneous porous media
smaller than the REV can exhibit significant anisotropy.
We have confirmed the hypothesis that various measures of this finite-size anisotropy
decrease with the system size $L$ as $L^{-\delta}$ with $\delta=-3/2$.
These measures include the angle between the driving force
and the resulting volumetric fluid flux in systems with periodic boundary conditions as well as the pressure gradient
induced in flows through solid wall channels.
By taking $\delta = -d/2$, this conclusion can be generalized to quasi one-dimensional ($d=1$)
and quasi two-dimensional ($d=2$) flows \cite{Koza09}.

Perhaps the most important observation is that the $L^{-\delta}$ scaling
holds for systems much smaller than the REV. This implies that
computer simulations of macroscopic systems can be performed using relatively small, ``mesoscopic'' models,
provided that the results will be extrapolated to macroscopic scales using appropriate scaling.
Moreover, since the $L^{-\delta}$ scaling can be theoretically related
to the central limit theorem \cite{Koza09}, a
porous medium can be regarded as consisting of essentially \emph{independent} volumes \emph{much smaller} than the REV.
This should facilitate not only numerical, but also theoretical investigation of transport in porous media.

Our results were obtained for a particular choice of a porous medium which was
modelled as a union of overlapping objects of the same size and shape. We expect that our conclusions
are quite general and are also valid for other classes of statistically uniform porous media,
e.g.\ for polidisperse porous media or for porous matrices generated
using different algorithms, e.g.\ by convoluting random fields with Gaussian kernels \cite{Hyman12}
or by using the data from the X-ray computed tomography \cite{Sugita12,Zhang00}.
However, this hypothesis needs to be verified independently.

Finally, our work confirms usability of the Sailfish library for investigation of flows in porous media.
While the results obtained with this software differ by several percent from the ones obtained
using a more traditional approach that is routinely used in computational fluid dynamics (FVM  and PISO),
both methods yield qualitatively the same results.
However, Sailfish, which is an open-source GPU implementation of the lattice Boltzmann method,
turns out faster, requires far less computer resources, and can handle much larger problems.

\section{Acknowledgments}

This work is supported by MNiSW grant No.\ N
519 437939 (ZK and MM). MJ acknowledges a scholarship from the TWING project
cofinanced by the European Social Fund.
This research was supported in part by PL-Grid Infrastructure.

\bibliographystyle{elsarticle-num-names}
%\bibliography{tort}

\end{document}